\documentstyle[11pt]{article}
\begin{document}
\vspace*{-2cm}

\large
ADP-92-192/T120

\vspace*{0.4cm}

\huge
\begin{center}
Shadowing in Deuterium.
\end{center}

\vspace*{0.4cm}

\large
\begin{center}
W.Melnitchouk and A.W.Thomas                            \\
Department of Physics and Mathematical Physics          \\
University of Adelaide                                  \\
Box 498 G.P.O., Adelaide, 5001, Australia
\end{center}

\vspace*{0.2cm}
\normalsize
\begin{center}
Abstract
\end{center}

\vspace*{0.2cm}

\hspace*{-0.5cm}
We calculate nuclear shadowing in lepton-deuteron deep inelastic scattering,
which arises from the double scattering of the virtual photon from both
nucleons in the deuteron.
The total correction to the deuteron structure function is found to be
$\stackrel{<}{\sim} 1\%$ at small $x$, but dependent on the model deuteron
wavefunction.
The resulting increase in the corrected neutron structure function is
$\sim 1-2\%$ for $x \simeq 0.004$, which leads to a $4-10\%$ decrease in the
value of the Gottfried sum obtained recently by the New Muon Collaboration.

\vspace*{8cm}


\newpage

\section{Introduction}

The quark structure of the nucleon is one of the most fundamental aspects
of hadron physics. Deep inelastic scattering (DIS) of leptons from hydrogen
has yielded a wealth of information on the deep inelastic structure of
the proton.
However, the absence of free neutron targets has forced one to use deuterium
in order to extract data on the neutron structure functions.
Traditionally in DIS on the deuteron, in which the proton and neutron
are held together very weakly, nuclear effects have been ignored,
and the total lepton-deuteron cross section assumed to be the sum of
the lepton-proton and lepton-neutron cross sections.
It is the deviation from this simple relation in the region of small
Bjorken $x$ ($x \stackrel{<}{\sim} 0.1$) which is known as shadowing.

Experimentally, a deviation from linearity has been observed \cite{emc} in the
so-called nuclear EMC effect for the ratio of DIS cross sections (or structure
functions) for lepton scattering from a nucleus and from deuterium.
A dramatic decrease in the nuclear structure function per nucleon in the
region of small $x$ confirmed earlier predictions \cite{nz75} that shadowing
should be present in DIS.
Furthermore, the shadowing was found to be only weakly dependent on $Q^{2}$.
The extraction of information about the difference between
nuclear structure functions and those for the free nucleon from
the observed nucleus/deuterium ratios is sensitive to
any nuclear effects in the deuteron.
Conclusions made about nucleon parton distributions based on
the nuclear/deuteron structure function ratios (eg. for the proton antiquark
distributions in the Drell-Yan process \cite{dryan}) at small $x$ may have
to be modified once shadowing is taken into account.

A precise knowledge of the neutron structure function, $F_{2n}$,
is essential for the determination
of the Gottfried sum rule, and the corresponding resolution of the question
of flavour symmetry violation in the proton sea.
It is necessary therefore to check for nuclear shadowing effects in
deuterium and include this correction in the extraction of $F_{2n}$ from
the deuteron structure function, $F_{2D}$.
Some recent estimates \cite{zol,nizo} have suggested a significant
amount of shadowing in deuterium (up to 4\%) for $x \stackrel{<}{\sim} 0.1$.
Other calculations \cite{bdkw} have predicted a less
dramatic effect ($\approx$ 2\%).

The cross section for lepton-deuteron DIS, fig.1, is related to the
forward $\gamma^{*} D$ scattering amplitude.
In the impulse approximation, fig.2, the virtual
photon interacts with one of the nucleons in the nucleus. The double
scattering diagram, fig.3, in which both nucleons participate in the
interaction, is the origin of the shadowing in a nucleus.

\section{Vector Meson Dominance}

\underline{Hadron-Deuteron Glauber Scattering}

Glauber theory \cite{glaub,grib} for hadron-deuteron scattering gives the
total $h D$ cross section as a sum of the $h N$ cross sections, and
a screening term arising from the double scattering of both nucleons:
\begin{eqnarray}
\sigma_{hD}
& = & 2 \sigma_{hN} + \delta \sigma_{hD}
\end{eqnarray}
where
\begin{eqnarray}
\delta \sigma_{hD}
& = & -\frac{ \sigma_{hN}^{2} }{ 8 \pi^{2} }
       \int d^{2}{\bf k}_{T}\ S_{D}({\bf k}^{2})                    \nonumber
\\
& = & -\frac{ \sigma_{hN}^{2} }{ 4 \pi }
       \int dk\ k\ S_{D}({\bf k}^{2}),                          \label{dsighD}
\end{eqnarray}
with $k \equiv |{\bf k}|$.
In deriving $\delta\sigma_{hD}$, the assumption is made that the
hadron-nucleon scattering amplitude, ${\cal F}_{hN}$,
is primarily imaginary, ${\rm Re} {\cal F}_{hN} \ll {\rm Im} {\cal F}_{hN}$,
and approximately independent of ${\bf k}^{2}$ for small ${\bf k}^{2}$.
(Contributions to $\delta \sigma_{hD}$ from large ${\bf k}^{2}$ will be
suppressed by the deuteron form factor $S_{D}({\bf k}^{2})$.)
Then from the forward double scattering amplitude \cite{aber}
\begin{eqnarray}
\delta {\cal F}_{hD}
& = & \frac{i}{2\pi |{\bf q}|}
      \int d^{2}{\bf k}_{T}\ S_{D}({\bf k}^{2})\
      {\cal F}_{hp}({\bf k}^{2})\ {\cal F}_{hn}({\bf k}^{2}),
\end{eqnarray}
where ${\bf q}$ is the momentum of the projectile,
eqn.(\ref{dsighD}) follows via the optical theorem:
$$
\sigma = \frac{ 4\pi }{ |{\bf q}| }\ {\rm Im} {\cal F}.
$$

\underline{$\gamma^{*} D$ Scattering}

Assuming that the Glauber formalism can be applied to $\gamma^{*} D$
scattering,
the shadowing correction to the $\gamma^{*} D$ cross section was originally
calculated
in terms of the vector meson dominance (VMD) model, where the virtual photon
dissociates into its hadronic components (vector mesons) before interacting
with the nucleon -- see fig.4.
In this model the shadowing cross section is given by \cite{vmd}
\begin{eqnarray}
\delta^{(V)} \sigma_{\gamma^{*} D}
& = & \sum_{v} \frac{ e^{2} }{ f_{v}^{2} }
      \frac{ 1 }{ (1 + Q^{2}/M_{v}^{2})^{2} } \delta\sigma_{vD}  \label{dsigV}
\end{eqnarray}
where $v = \rho^{0}, \omega, \phi$, and the photon--vector meson coupling
constants are \cite{pdat}
\begin{eqnarray}
\frac{ f_{v}^{2} }{ 4 \pi }
& = & \frac{ \alpha^{2} M_{v} }{ 3\ \Gamma_{v\rightarrow e^{+}e^{-}} }
\end{eqnarray}
(equal to 2.28, 26.14, 14.91 for $\rho^{0}, \omega$ and $\phi$, respectively
\footnote{Note that the fine structure constant evaluated at
$Q^{2}={\cal O}(1$GeV$^{2})$ is $\alpha = e^{2}/4\pi \approx 1/130$,
although the error introduced by this is probably less than that associated
with using $f_{v}^{2}$, which is obtained from the decay of meson $v$ with
time-like $Q^{2}$, for the coupling to a photon with space-like $Q^{2}$.}).
Writing (\ref{dsigV}) in terms of the deuteron structure function,
$F_{2D}$
\footnote{In terms of the total cross section for the photo-absorption
of virtual photons on an unpolarised deuteron, $\sigma_{\gamma^{*} D}$, the
deuteron structure function is
$$ W_{2D}
= \frac{K}{4 \pi^{2} \alpha} \frac{Q^{2}}{Q^{2}+\nu^{2}}
\sigma_{\gamma^{*} D} $$
where $K = \sqrt{\nu^{2} + Q^{2}}$ is the flux of incoming virtual photons
(in the Gilman convention), so that in the Bjorken limit
$$ F_{2D} = \frac{Q^{2}}{4 \pi^{2} \alpha} \sigma_{\gamma^{*} D}.$$},
we have
\begin{eqnarray}
\delta^{(V)} F_{2D}(x)
& = & \frac{ Q^{2} }{ \pi }
\sum_{v} \frac{ \delta\sigma_{vD} }{ f_{v}^{2} (1 + Q^{2}/M_{v}^{2})^{2} },
                                                                \label{dFV}
\end{eqnarray}
where now
\begin{eqnarray}
\delta \sigma_{vD}
& = & -\frac{ \sigma_{vN}^{2} }{ 8 \pi^{2} }
       \int d^{2}{\bf k}_{T}\ S_{D}({\bf k}^{2}).               \label{dsigVD}
\end{eqnarray}
The total vector meson--nucleon cross sections, $\sigma_{vN}$, are related to
the total $\pi N$ and $K N$ cross sections via the quark model,
and are set to 24 mb for $v = \rho^{0}$ and $\omega$,
and 14.5 mb for $v = \phi$ (see \cite{vmd,vNqm}).
The deuteron form factor $S_{D}({\bf k}^{2})$ is given by the electric monopole
body form factor \cite{CE}
\begin{eqnarray}
S_{D}({\bf k}^{2})
& = & \int_{0}^{\infty} dr \left( u^{2}(r) + w^{2}(r) \right)
                           j_{0}(k r),                          \label{SD}
\end{eqnarray}
where $u(r), w(r)$ are the $S,D$-wave deuteron wavefunctions, normalised s.t.
$\int dr\ (u^{2}(r) + w^{2}(r)) = 1$, and where $j_{0}$ is the
spherical Bessel function.
The square of the 3-momentum transfer to the interacting nucleon is
${\bf k}^{2} = {\bf k}_{T}^{2} + k_{L}^{2}$, where
$k_{L}^{2} = m_{N}^{2} x^{2} (1 + M_{v}^{2}/Q^{2})^{2}$,
and $x = Q^{2} / 2 p \cdot q$.

 From eqn.(\ref{dFV}) it can be seen that the VMD shadowing correction to
the deuteron structure function decreases as $1/Q^{2}$ for $Q^{2}
\rightarrow \infty$.

At $Q^{2} = 4$GeV$^{2}$ the VMD model shadowing predictions are given in
fig.5 for deuteron form factors obtained from several different $NN$ potential
models. By far the largest contribution ($\approx 80\%$) to
$\delta^{(V)} F_{2D}$ comes from the $\rho^{0}$ meson.
The magnitude of $\delta^{(V)} F_{2D}(x)$ decreases with $x$ because the
lower limit of the $k$-integration in eqn.(\ref{dsigVD}), $k_{min} = k_{L}$,
is an increasing function of $x$, and the integrand peaks at small values of
$k\ (\approx 0.7$\ fm$^{-1}$).
The model dependence arises from differences in the
large-$k\ (\stackrel{>}{\sim} 2$\ fm$^{-1})$ behaviour of the form factor,
fig.6,
which itself is determined by the small-$r$ behaviour of $u(r), w(r)$.
All of the deuteron wavefunctions obtained from realistic $NN$ potential models
(namely Paris \cite{paris}, Bonn (OBEPQ) \cite{bonn} and Bochum \cite{boch})
produce a trough in $k S_{D}({\bf k}^{2})$ at $k \approx 3.5$\ fm$^{-1}$
(because the Bessel function is negative at large $kr$), and a rapid fall-off
with $k$ for $k \stackrel{>}{\sim} 6$\ fm$^{-1}$.
Also shown is the model of Franco and Varma \cite{fran}, which was used in
\cite{zol,nizo}, for which the form factor, parameterised by a sum of
Gaussians, has no large-$k$ tail at all.
The form factor with the Paris wavefunction, which has the `deepest' trough,
leads to $\delta^{(V)} F_{2D}$ which is $\approx 25\%$ smaller for
$x \stackrel{<}{\sim} 0.01$ than with the Franco and Varma form factor.
The trough is also responsible for the antishadowing
in the region $x \stackrel{>}{\sim} 0.2$.

\section{Diffractive Scattering from Partons}

At low $Q^{2}$, it is most natural to evaluate the $\gamma^{*} D$ shadowing in
terms
of the VMD model. At higher energies a parton picture may be more relevant.
An alternative description of the double interaction mechanism in fig.3
in the high energy limit is in terms of Pomeron (${\cal P}$) exchange, fig.7.
If the momentum transfer between the photon and nucleon is small, the nucleon
will most likely remain intact, in which case there will only be exchange of
vacuum quantum numbers.
Although there is as yet no QCD-based derivation of the properties of the
reactions described by Pomeron exchange (eg. constant hadronic cross sections),
there have been suggestions \cite{low,nus,glr} that the Pomeron represents
a system of gluons.
(In ref.\cite{low} hadron-hadron scattering is modelled in terms of gluon
exchange between MIT bags, while in ref.\cite{glr} gluon-ladder techniques
are used to calculate deep inelastic structure functions of hadrons at
low $x$.)

In fig.7 the virtual photon probes the parton structure of the Pomeron,
which is parameterised by the Pomeron structure function $F_{2{\cal P}}$
\cite{ingel,dola} (defined in terms of the cross section for
$\gamma^{*}$--Pomeron
diffractive scattering):
\begin{eqnarray}
F_{2{\cal P}}(x) & \equiv & \frac{ Q^{2} }{ 4\pi^{2}\alpha }
\sigma_{\gamma^{*}{\cal P}}.
\end{eqnarray}
The contribution to the $F_{2D}$ structure function from multiple diffractive
scattering with ${\cal P}$ exchange can be written as a convolution of an
exchange-${\cal P}$ distribution function, $f_{{\cal P}}(y)$, with the
${\cal P}$ structure
function:
\begin{eqnarray}
\delta^{({\cal P})} F_{2D}(x)
& = & \int_{y_{min}}^{2}\ dy\ f_{{\cal P}}(y)\ F_{2{\cal P}}(x_{{\cal P}})
\label{dFP}
\end{eqnarray}
where
\begin{eqnarray}
f_{{\cal P}}(y) & = & - \frac{\sigma_{pp}}{8 \pi^{2}}\ \frac{1}{y}
                 \int d^{2}{\bf k}_{T}\ S_{D}({\bf k}^{2})
\end{eqnarray}
is expressed as a function of the momentum fraction of the nucleon carried
by the Pomeron, $y = k \cdot q / p \cdot q
                   = x (1 + M_{X}^{2}/Q^{2})
             \approx M_{X}^{2}/s$\ \
($M_{X}^{2} = p_{X}^{2}, s = (p + q)^{2}$), and we define $x_{{\cal P}}
\equiv x/y$.
Fig.8 illustrates the $y$-dependence of $f_{{\cal P}}(y)$, including the $1/y$
divergence for $y \rightarrow 0$. The rapid fall off with $y$ is testament
to the very small contribution coming from the large-$y$ region.

In formulating a complete description of shadowing which includes more than
one mechanism care must be taken to avoid possible double counting.
Because of this concern some authors \cite{bdkw} have restricted the
Pomeron exchange process to the region of $M_{X}^{2}$ above the highest
mass of the vector mesons contributing to the VMD process:
$M_{X}^{2} \geq M_{X_{0}}^{2} \simeq 1.5$GeV$^{2}$, and
consequently have taken the lower bound on the integral in eqn.(\ref{dFP})
to be $y_{min} = x (1 + M_{X_{0}}^{2}/Q^{2})$.
The VMD contribution, which is essentially a higher twist ($1/Q^{2}$) effect,
may compete with that part of the diagram in fig.7 which contains low-$M_{X}$
single particle intermediate states.
By keeping only the leading twist piece of the structure function
$F_{2{\cal P}}$,
we can exclude this contribution since it involves extra factors of
$1/Q^{2}$ from the electromagnetic form factors.
Nevertheless, we have tested the sensitivity of our numerical results to the
cut-off procedure by varying $M_{X_{0}}^{2}$ from 0 to 2 GeV$^{2}$.
For low $x$ we find a difference over this range of only some 5\% of the
total ${\cal P}$ exchange contribution to $F_{2D}$.
For larger $Q^{2}$ the separation into separate $M_{X}$ regions becomes
irrelevant since $y_{min} \rightarrow x$ in the Bjorken limit.

For the Pomeron structure function we include contributions from the
quark-antiquark box diagram, fig.9a, and from the triple Pomeron
interaction, fig.9b (see refs.\cite{kwbd,niza}):
\begin{eqnarray}
F_{2{\cal P}}(x_{{\cal P}})
& = & F_{2{\cal P}}^{(box)}(x_{{\cal P}})
  + F_{2{\cal P}}^{(3{\cal P})}(x_{{\cal P}})
\end{eqnarray}
normalised such that
\begin{eqnarray}
F_{2{\cal P}} & = & \left. \left( \frac{ 16 \pi y }{ \sigma_{pp} } \right)
                             \frac{ d^{2} F_{2}^{diff} }{ dt\ dy }
               \right|_{t=0},
\end{eqnarray}
where $t \approx -{\bf k}^{2}$, and $F_{2}^{diff}$ is the diffractive structure
function, describing semi-inclusive diffractive lepton-nucleon DIS, in which
the recoil nucleon and the hadronic state $X$ are separated by a large
rapidity \cite{dola}.

The Pomeron structure function arising from the quark box diagram,
$F_{2{\cal P}}^{(box)}$, has been calculated by Donnachie and Landshoff
\cite{dola}:
\begin{eqnarray}
F_{2{\cal P}}^{(box)}(x_{{\cal P}})
& = & \frac{ ( 12 \Sigma_{q^{2}}\ N_{sea})\ \beta_{0}^{2} }{ \sigma_{pp} }\
      x_{{\cal P}} (1-x_{{\cal P}}).                              \label{Fbox}
\end{eqnarray}
The quark---Pomeron coupling constant is $\beta_{0}^{2} = 3.4 $GeV$^{-2}$
\cite{dl84}, and we assume the same strength for $u,d$ quark and
antiquark---Pomeron couplings, but a weaker coupling to the strange quarks:
$\Sigma_{q^{2}} = (10 + 2 \lambda_{s})/9$ with $\lambda_{s} \simeq 0.5$.
According to the Particle Data Group \cite{pdat}, the proton-proton total
cross section $\sigma_{pp}$ is approximately 40 mb.
The parameter $N_{sea}$ is determined by the $x \rightarrow 0$ behaviour of the
nucleon sea distribution,
$x q_{sea}(x \rightarrow 0) \rightarrow N_{sea} x^{a}$.
Recent parameterisations of world DIS, Drell-Yan and prompt photon data
\cite{hmrs,ow,mt} give $N_{sea} \simeq 0.15$, and $a$ approximately 0.
Note that the overall normalisation of the r.h.s. of eqn.(\ref{Fbox}) is
slightly smaller than in \cite{bdkw} due to our smaller sea parameter
$N_{sea}$ (cf. $N_{sea}=0.17$ in \cite{bdkw}) and suppression of
strange---Pomeron couplings.
More recently, Nikolaev and Zakharov \cite{niza} have calculated
the box diagram contribution to $F_{2{\cal P}}$,
based on a perturbative QCD analysis of $q \bar{q}$ fluctuations
of the virtual photon.
The $x_{{\cal P}}$ dependence of their $F_{2{\cal P}}^{(box)}$ parameterisation
is the same as that in eqn.(\ref{Fbox}):
$M_{X}^{2} / (Q^{2} + M_{X}^{2})^{3}$
(since $Q^{2} + M_{X}^{2} = Q^{2} / x_{{\cal P}}$\ from
the definition of $x_{{\cal P}}$), providing the same normalisation
is used (the normalisations in \cite{dola} and \cite{niza,bgnpz}
differ by an overall factor $1-x_{{\cal P}}$).

The triple Pomeron part of the ${\cal P}$ structure function,
\begin{eqnarray}
F_{2{\cal P}}^{(3{\cal P})}(x_{{\cal P}})
& = & \frac{ 16 \pi }{ \sigma_{pp} }
      \left[ \frac{ y }{ \sigma_{hp} }
         \left. \frac{ d^{2}\sigma_{hp\rightarrow hX} }{ dt dy }
         \right|_{t=0}
      \right] F_{2N}^{sea}(x_{{\cal P}},Q^{2})                   \label{F3P}
\end{eqnarray}
follows from
\begin{eqnarray}
\frac{ 1 }{ F_{2N}^{sea} } \left. \frac{ d^{2}F_{2}^{diff} }{ dtdy }
                           \right|_{t=0}
& = &
\frac{ 1 }{ \sigma_{hp} } \left. \frac{ d^{2}\sigma_{hp \rightarrow hX} }
                                      { dtdy }
                          \right|_{t=0}
\end{eqnarray}
and the Regge theory expression for the diffractive differential cross section
\cite{goul}
\begin{eqnarray}
\frac{ d^{2}\sigma_{hp \rightarrow hX} }{ dtdy }
& = &
\frac{ \beta_{h{\cal P}}(t)\ \beta_{p{\cal P}}^{2}(t)\ g_{3{\cal P}}(t) }
     { 16\pi }
y^{1 - 2 \alpha_{{\cal P}}(t)}
\end{eqnarray}
where $\alpha_{{\cal P}}(t) \approx 1 + 0.25 t$.
In the Regge model the total $h p$ cross
section is also given in terms of the hadron---Pomeron couplings,
$\beta_{h{\cal P}}$: $\sigma_{hp} = \beta_{h{\cal P}}(0) \beta_{p{\cal P}}(0)$.
It is then evident that the combination
\begin{eqnarray}
\frac{ 1 }{ \sigma_{hp} } \left. \frac{ d^{2}\sigma_{hp \rightarrow hX} }
                                      { dtdy }
                          \right|_{t=0}
& = & \frac{ \beta_{p{\cal P}}(0)\ g_{3{\cal P}}(0) }{ 16 \pi y } \label{reg3P}
\end{eqnarray}
is independent of hadron $h$. From experiments
on the diffractive dissociation of $\pi^{\pm}, K^{\pm}, p$
and $\bar{p}$ on hydrogen, the triple Pomeron coupling constant was found to
be $g_{3{\cal P}}(0) \simeq 0.364$ mb$^{1/2}$ \cite{3Pexp}, independent of $t$,
and indeed of the hadron type $h$.

For the sea part of the nucleon structure function,
$F_{2N}^{sea} = 5 x ( u_{s} + \bar{u} + d_{s} + \bar{d}
                    + 2(s + \bar{s})/5 ) / 18$,
we use recent parameterisations of the data at $Q^{2} = 4 $GeV$^{2}$
\cite{ow,mt}.
In the calculation of ref.\cite{bdkw}, a constant value of 0.3 was used for
$F_{2N}^{sea}$ together with an empirical low-$Q^{2}$ dependence \cite{dola}.
With the above triple Pomeron coupling constant,
eqn.(\ref{F3P}) gives a $3{\cal P}$ component which is about 40\% smaller
than that obtained in \cite{zol}. However, this is not very significant
for the total Pomeron structure function, since $F_{2{\cal P}}^{(3{\cal P})}$
is
very much smaller than the quark-antiquark `box' contribution,
$F_{2{\cal P}}^{(box)}$, fig.10.

The scaling behaviour of the ${\cal P}$-exchange mechanism is determined by the
scaling behaviour of the ${\cal P}$ structure function, and from
eqns.(\ref{Fbox})---(\ref{reg3P}) it is clear that $\delta^{({\cal P})} F_{2D}$
will scale as $Q^{2} \rightarrow \infty$.
At $Q^{2} = 4$GeV$^{2}$, fig.11a shows the individual `box' and 3${\cal P}$
contributions to $\delta^{({\cal P})} F_{2D}$, with the deuteron form factor
obtained from the Bochum wavefunction.
The dependence of $\delta^{(P)} F_{2D}$ on $S_{D}({\bf k}^{2})$
is illustrated in fig.11b. Again, as in the case of the VMD
model, the large-$k$ negative tail of the form factor produces a large
(some 30-40\%) difference between different models for
$x \stackrel{<}{\sim} 0.05$.
For $x \stackrel{>}{\sim} 0.2$ the presence or absence of antishadowing will be
determined by the model deuteron wavefunction.

\section{Shadowing by Mesons}

Another potential source of shadowing arising from the double scattering
mechanism is one which involves the exchange of mesons, fig.12.
It has previously been suggested \cite{kap} that this
leads to substantial antishadowing corrections to $F_{2D}(x)$.
The total contribution to the deuteron structure function from meson exchange
is written
\begin{eqnarray}
\delta^{(M)} F_{2D}(x)
& = & \sum_{\mu}
      \int_{x}^{m_{D}/m_{N}} dy\ f_{\mu}(y)\ F_{2\mu}(x_{\mu}),  \label{dFM}
\end{eqnarray}
where $\mu = \pi,\rho,\omega,\sigma$ and
$y = k \cdot q / p \cdot q = (k_{0} + k_{L})/m_{N}$ and $x_{\mu} = x/y$.
For the virtual meson structure function, $F_{2\mu}$, we take the
parameterisation of the (real) pion structure function from Drell-Yan
production \cite{F2pi}.
The exchange-meson distribution functions $f_{\mu}(y)$ are obtained from the
non-relativistic reduction of the nucleon---meson interaction:
\begin{equation}
\hspace*{-0.3cm}
f_{\mu}(y) = 4 c_{\mu}\ m_{N}
\int \frac{ d^{3}{\bf p}\ d^{3}{\bf p'} }{ (2\pi)^{3} }
\frac{ {\cal F}^{2}_{\mu NN}(k^{2}) }{ (k^{2} - m_{\mu}^{2})^{2} }
\ y\ \left\{ \frac{1}{3} \sum_{J_{z}}
             \Psi^{\dagger}({\bf p},J_{z})\ {\cal V}_{\mu NN}\
             \Psi({\bf p'},J_{z})\
     \right\}
     \delta\left( y - \frac{k_{0} + k_{L}}{m_{N}} \right).        \label{fyM}
\end{equation}
The deuteron wavefunction is defined by
\begin{eqnarray}
\Psi({\bf p},J_{z})
& = & \frac{1}{\sqrt{4\pi}}
      \left( u(p) - w(p) \frac{ S_{12}(\hat{p}) }{ \sqrt{8} }
      \right)\ \chi_{1}^{J_{z}},
\end{eqnarray}
where $u(p)$ and $w(p)$ are its $S$ and $D$-wave components,
normalised so that $\int dp\ {\bf p}^{2}\ (u^{2}(p) + w^{2}(p)) = 1$,
with $\hat{p} \equiv {\bf p}/p$ and $p \equiv |{{\bf p}}|$,
and $S_{12}$ is the tensor operator:
$S_{12}(\hat{p}) = 3\ \sigma_{1} \cdot \hat{p}\ \sigma_{1} \cdot \hat{p}
                 - \sigma_{1} \cdot \sigma_{2}$.
The deuteron spin wavefunction is denoted by $\chi_{1}^{J_{z}}$,
where $J_{z}$ is the total angular momentum projection.
In eqn.(\ref{fyM}), $k^{2} = k_{0}^{2} - {\bf k}^{2}$, where
$k_{0} = m_{D} - \sqrt{m_{N}^{2} + {\bf p}^{2}} - \sqrt{m_{N}^{2}
       + {\bf p'}^{2}}$
is the energy of the off-shell meson, and ${\bf k} = {\bf p} - {\bf p'}$ is its
3-momentum.

The nucleon---meson interactions are given by \cite{bonn}
\begin{eqnarray}
{\cal V}_{\pi NN}
& = & - \frac{ f_{\pi NN}^{2} }{ m_{\pi}^{2} }\
      \sigma_{1} \cdot {\bf k}\ \sigma_{2} \cdot {\bf k}\        \\
{\cal V}_{\rho NN}
& = & g_{\rho NN}^{2}
\left[ 1 + \frac{ 3 {\bf q}^{2} }{ 2 m_{N}^{2} }
         - \frac{ {\bf k}^{2} }{ 8 m_{N}^{2} }
         - \sigma_{1} \cdot \sigma_{2} \frac{ {\bf k}^{2} }{ 4 m_{N}^{2} }
     + \frac{ \sigma_{1} \cdot {\bf k}\ \sigma_{2} \cdot {\bf k} }
            { 4 m_{N}^{2} }
\right]                                                          \nonumber \\
& + & \frac{ g_{\rho NN} f_{\rho NN} }{ 2 m_{N} }
\left[ - \frac{ {\bf k}^{2} }{ m_{N} }
       - \sigma_{1} \cdot \sigma_{2} \frac{ {\bf k}^{2} }{ m_{N} }
       + \frac{ \sigma_{1} \cdot {\bf k}\ \sigma_{2} \cdot {\bf k} }
              { m_{N} }
\right]                                                                 \\
& + & \frac{ f_{\rho NN}^{2} }{ 4 m_{N}^{2} }
\left[ - \sigma_{1} \cdot \sigma_{2}\ {\bf k}^{2}
       + \sigma_{1} \cdot {\bf k}\ \sigma_{2} \cdot {\bf k}
\right]                                                          \nonumber \\
{\cal V}_{\omega NN}
& = & g_{\omega NN}^{2}
\left[ 1 + \frac{ 3 {\bf q}^{2} }{ 2 m_{N}^{2} }
     - \frac{ {\bf k}^{2} }{ 8 m_{N}^{2} }
     - \sigma_{1} \cdot \sigma_{2} \frac{ {\bf k}^{2} }{ 4 m_{N}^{2} }
     + \frac{ \sigma_{1} \cdot {\bf k}\ \sigma_{2} \cdot {\bf k} }
            { 4 m_{N}^{2} }
\right]                                                                 \\
{\cal V}_{\sigma NN}
& = & - g_{\sigma NN}^{2}
\left[ 1 - \frac{ {\bf q}^{2} }{ 2 m_{N}^{2} }
         + \frac{ {\bf k}^{2} }{ 8 m_{N}^{2} }
\right],
\end{eqnarray}
where ${\bf q} = \frac{1}{2} ({\bf p} + {\bf p'})$.
Terms proportional to ${\bf S} \cdot {\bf k} \times {\bf q}$,
where ${\bf S} = \sigma_{1} + \sigma_{2}$, are omitted as they do not
contribute to $f_{\mu}(y)$.

Evaluation of eqn.(\ref{fyM}) requires the identities:
\begin{eqnarray}
\frac{1}{3} \sum_{J_{z}} \Psi^{\dagger}({\bf p},J_{z})\ \Psi({\bf p'},J_{z})\
& = & \frac{1}{4\pi} \left[ u(p)\ u(p') + w(p)\ w(p')\
                            P_{2}(\cos\theta)\ P_{2}(\cos\theta')
                     \right]                                    \nonumber\\
& + & \phi\ {\rm dependent\ terms}                    \label{ident1}\\
& = & \frac{1}{3} \sum_{J_{z}} \Psi^{\dagger}({\bf p},J_{z})\
      \sigma_{1} \cdot \sigma_{2}\ \Psi({\bf p'},J_{z})             \nonumber
\end{eqnarray}
\begin{eqnarray}
\frac{1}{3} \sum_{J_{z}} \Psi^{\dagger}({\bf p},J_{z})\
\sigma_{1}\cdot {\bf k}\ \sigma_{2}\cdot {\bf k}\ \Psi({\bf p'},J_{z})
& = & \frac{1}{4\pi}
\left\{ \frac{1}{3} \left[ {\bf k}^{2} - 2\ p\ p'\ \sin\theta\ \sin\theta'\
                    \right] u(p)\ u(p')
\right.                                                         \nonumber\\
& - & \frac{1}{\sqrt{2}}
\left[ 4\ p\ p'\ \cos\theta\ \cos\theta'\ \sin^{2}\theta'\
    +\ 4\ {\bf p'}^{2} \cos^{2}\theta'\ \sin^{2}\theta'\
\right.                                                         \nonumber\\
&   & -\ \frac{2}{3}\ ({\bf p}^{2} + {\bf p'}^{2})\ P_{2}(\cos\theta')\
      + 2\ ({\bf p}^{2} \cos^{2}\theta
 + {\bf p'}^{2} \cos^{2}\theta')\ P_{2}(\cos\theta')            \nonumber\\
&   & +\ \left. \frac{8}{3}\ p\ p'\ \cos\theta\ \cos\theta'\ P_{2}(\cos\theta')
         \right]\ u(p)\ w(p')                                   \nonumber\\
& - & \frac{1}{\sqrt{2}}
\left[ 4\ p'\ p\ \cos\theta'\ \cos\theta\ \sin^{2}\theta
  +\ 4\ {\bf p}^{2} \cos^{2}\theta\ \sin^{2}\theta
\right.                                                         \nonumber\\
&   & -\ \frac{2}{3}\ ({\bf p'}^{2} + {\bf p}^{2})\ P_{2}(\cos\theta)\
      +\ 2\ ({\bf p'}^{2} \cos^{2}\theta'
    + {\bf p}^{2} \cos^{2}\theta)\ P_{2}(\cos\theta)            \nonumber\\
&   & +\ \left. \frac{8}{3}\ p'\ p\ \cos\theta'\ \cos\theta\ P_{2}(\cos\theta)
         \right]\ w(p)\ u(p')                                   \nonumber\\
& - & \frac{1}{3}
\left[ (p\ \cos\theta + p'\ \cos\theta')^{2}\ P_{2}(\cos\theta)\
                                              P_{2}(\cos\theta')
\right.                                                         \nonumber\\
&   & -\ 2\ ({\bf p}^{2} \sin^{2}\theta
   + {\bf p'}^{2} \sin^{2}\theta')\ P_{2}(\cos\theta)\ P_{2}(\cos\theta')\
                                                                \nonumber\\
&   & +\ 3\ ({\bf p}^{2} \cos^{2}\theta\ \sin^{2}\theta
+ p\ p'\ \cos\theta\ \cos\theta'\ \sin^{2}\theta)\ P_{2}(\cos\theta')
\nonumber\\
&   & +\ 3\ ({\bf p'}^{2} \cos^{2}\theta'\ \sin^{2}\theta'
+ p'\ p\ \cos\theta'\ \cos\theta\ \sin^{2}\theta')\ P_{2}(\cos\theta)
\nonumber\\
&   &
\left. \left.
+\ \frac{9}{2}\ p\ p'\ \cos\theta\ \cos\theta'\ \sin^{2}\theta\ \sin^{2}\theta'
       \right]\ w(p)\ w(p')
\right\}                                            \label{ident2}\\
& + & \phi\ {\rm dependent\ terms}.                             \nonumber
\end{eqnarray}
The terms in eqns.(\ref{ident1})-(\ref{ident2}) which depend on
the azimuthal angle ($\phi$) vanish after integration.
The factors $c_{M}$ are due to isospin:
$c_{\pi} = c_{\rho} = 3, c_{\omega} = c_{\sigma} = -1$.
The $\mu NN$ vertex form factors ${\cal F}_{\mu NN}(k^{2})$ are parameterised
by a dipole form
\begin{eqnarray}
{\cal F}_{\mu NN}(k^{2})
& = & \left( \frac{ \Lambda_{\mu}^{2} - m_{\mu}^{2} }
                  { \Lambda_{\mu}^{2} - k^{2} }
      \right)^{2},
\end{eqnarray}
with the high-momentum cut-offs $\Lambda_{\mu}$ ranging from
$\sim 1$GeV in models with soft form factors \cite{ht90,boch} to
$\sim 1.7-2$GeV when hard form factors are employed \cite{bonn}.

Fig.13 shows the individual meson exchange contributions to
$\delta^{(M)} F_{2D}$, for the wavefunction of the Bonn model,
and with a universal dipole cut-off of $\Lambda_{\mu} = 1.7$GeV.
As could be expected, pion exchange is the dominant process.
We also include the fictitious $\sigma$ meson,
but with a mass ($\approx$ 800 MeV) that is larger than that
used to represent $2 \pi$ exchange in $N N$ scattering.
Both of these produce antishadowing for small $x$.
The exchange of vector mesons ($\rho$, $\omega$) cancels some of this
antishadowing, although the magnitude of these contributions is smaller.
In fact, for $\Lambda_{\mu} \stackrel{<}{\sim} 1.3$ GeV all contributions other
than that of the pion are totally negligible.

Fig.14 shows the dependence of the total $\delta^{(M)} F_{2D}$ on
$\Lambda_{\mu}$ for the Bonn model wavefunction. There
is approximately a factor of 2 difference between the amount of shadowing
with soft ($\Lambda_{\mu} = 1$GeV, lower dotted line) and hard
($\Lambda_{\mu} = 1.7$GeV, upper dotted line) form factors.
In lepton-nucleon DIS it is well known \cite{t83} that the meson cloud of
the nucleon, with a hard $\mu NN$ form factor, gives nucleon sea
distributions that are several times larger than the empirical ones.
In fact, to be consistent with the lepton-nucleon DIS data
$\Lambda_{\mu}$ must be $\stackrel{<}{\sim} 0.8-0.9$GeV.
We also consider the effect of the model momentum-space
deuteron wavefunction on $\delta^{(M)} F_{2D}$.
Although the model wavefunctions differ substantially at large momenta
($p \stackrel{>}{\sim} 2\ $fm$^{-1}$), this variation will be largely
suppressed by the
$\mu NN$ form factor.
The Bochum and Paris wavefunctions are generally larger than the Bonn
wavefunction, and this is reflected in a larger $\delta^{(M)} F_{2D}$.

We also comment here on the issue raised in the previous section, namely
double counting, this time between the meson exchange and the other
mechanisms.
It should be clear that since the ${\cal P}$ contribution involves the exchange
of vacuum quantum numbers, there will be no interference between this and
the exchange of pseudoscalar pions or vector mesons.
The scalar $\sigma$ meson, introduced as an effective description of
two-pion $N \Delta$ excitations, does not correspond to actual exchange
of a spin 0 particle.
By restricting the meson structure function to only the leading twist
component (our $F_{2M}$ is determined at $Q^{2} = 25$ GeV$^{2}$ where
this assumption is reasonable) we may
view the VMD process as a description of higher twist effects.
Still, imposing any low-$M_{X}$ cut on the meson exchange term has
numerically insignificant consequences,
largely because $F_{2\mu}(x/y) \rightarrow 0$ as $y \rightarrow x$.

\section{Combined Shadowing Effects and the Gottfried Sum Rule}

The total deuteron structure function is defined by
\begin{eqnarray}
F_{2D}(x) & = & F_{2p}(x) + F_{2n}(x) + \delta F_{2D}(x),
\end{eqnarray}
where the shadowing correction is a sum of the VMD, Pomeron and meson exchange
contributions:
\begin{eqnarray}
\delta F_{2D}(x) & = & \delta^{(V)} F_{2D}(x)
                   +   \delta^{({\cal P})} F_{2D}(x)
                   +   \delta^{(M)} F_{2D}(x).
\end{eqnarray}
In fig.15 we compare the contributions to $\delta F_{2D}(x)$ from
the three mechanisms considered.
For $x \stackrel{<}{\sim} 0.1$ the magnitude of the (negative) Pomeron/VMD
shadowing
is larger than the (positive) meson-exchange contribution, so that the
total $\delta F_{2D}$ is negative. The fact that shadowing is present in
this region of $x$ does not depend on the model deuteron wavefunction.
For larger $x$ ($\approx 0.1-0.2$) there is a small amount of antishadowing,
which is due mainly to the VMD contribution.
The dependence of the total shadowing correction on the deuteron wavefunction
and on the $\mu NN$ form factor is shown in fig.16 for $Q^{2} = 4$ GeV$^{2}$.
We point out that the magnitude of $\delta F_{2D}(x)$ is about 4 times smaller
than that obtained in reference \cite{zol}, and about 2 times smaller compared
with the result of reference \cite{bdkw}.
The most important reasons for our smaller results are the inclusion of meson
exchange contributions which produce antishadowing at small $x$, and the
use of realistic deuteron wavefunctions which lead to smaller ${\cal P}$
 exchange
and VMD contributions.

Recently the New Muon Collaboration (NMC) has measured $F_{2p}$ and $F_{2D}$
\cite{nmc90,nmc} down to very small values of $x$ ($=x_{min}=0.004$).
The neutron structure function was then extracted from $F_{2D}$ in order
to test the Gottfried sum rule \cite{got}.
However, by assuming that
\begin{eqnarray}
\frac{ F_{2D}\ ( 1 - (F_{2D}/F_{2p} - 1) ) }
     {         ( 1 + (F_{2D}/F_{2p} - 1) ) }
& = & 2 F_{2p} - F_{2D}\ \ \equiv\ (F_{2p} - F_{2n})_{NMC}
\end{eqnarray}
the NMC ignored any nuclear shadowing effects in D which may alter the
$F_{2n}$ values.
The actual difference between the $p$ and $n$ structure functions
should be
\begin{eqnarray}
F_{2p} - F_{2n}
& = & (F_{2p} - F_{2n})_{NMC} + \delta F_{2D},
\end{eqnarray}
and this is shown in fig.17.
The dotted line is a best fit to the NMC data, and
includes the small-$x$ extrapolation used in \cite{nmc}:
\begin{eqnarray}
F_{2p}(x) - F_{2n}(x)
&  \stackrel{x \rightarrow 0}{\longrightarrow}  & \alpha\ x^{\beta}
\end{eqnarray}
with $\alpha = 0.21, \beta = 0.62$.
The other curves include the shadowing corrections to the NMC data
parameterisation.
It is not clear whether $F_{2p} - F_{2n}$ will become negative
at $x \stackrel{<}{\sim} 0.004$, and it will be interesting to see whether
this cross-over
occurs when additional data at smaller $x$ become available.

The Gottfried integral
\begin{eqnarray}
S_{G}(x,1)
& = & \int_{x}^{1} dx'\ \frac{ F_{2p}(x') - F_{2n}(x') }{ x' } \\ \nonumber
& = & \int_{x'=x}^{1} d\ (\log x')\ (F_{2p}(x') - F_{2n}(x'))
\end{eqnarray}
is given in fig.18 for $x$ down to 0.004.
In the naive quark model, $S_{G}(0,1) = 1/3$. Ignoring nuclear effects, the
NMC obtained $S_{G}(x_{min},1) = 0.229$. From the unmeasured region
($x < 0.004$), using the above extrapolation, the contribution was
found to be $S_{G}(0,x_{min}) = (\alpha/\beta)\ x_{min}^{\beta} = 0.011$.
With the conventional Regge theory assumption that $\beta = 0.5$,
$S_{G}(0,x_{min})$ would be 0.014.
In table 1 we give the values of $S_{G}$ including shadowing corrections,
and also the $x < x_{min}$ extrapolation parameters.
For simplicity we take $\beta = 0.5$, and adjust $\alpha$ to achieve a smooth
transition between the $x > x_{min}$ and $x < x_{min}$ regions.
The overall correction to the NMC value for $S_{G}(0,1)$ is found to be
between --0.010 and --0.026.
This is to be compared with --0.07 to --0.088 obtained in
\cite{zol,nizo,bgnpz}.

\begin{table}
\begin{tabular}{||l||lr|l|l|l||}
\hline\hline
Model                           & $\alpha$ & $\beta$ & $S_{G}(0,x_{min})$
& $S_{G}(x_{min},1)$ & $S_{G}(0,1)$       \\ \hline\hline
NMC \cite{nmc}                  & 0.21     & 0.62    & 0.011
& 0.229              &  0.240 $\pm$ 0.016 \\
                                & 0.109    & 0.5     & 0.014
&                    &  0.243             \\ \hline
Bochum ($\Lambda_{\mu}=1.3$GeV) & 0.043    & 0.5     & 0.005
& 0.222              &  0.227             \\ \hline
Paris ($\Lambda_{\mu}=1.3$GeV)  & 0.052    & 0.5     & 0.007
& 0.224              &  0.230             \\ \hline
Bonn ($\Lambda_{\mu}=1.3$GeV)   & 0.011    & 0.5     & 0.001
& 0.215              &  0.217             \\ \hline
Bonn ($\Lambda_{\mu}=1.0$GeV)   & 0.002    & 0.5     & 0.000
& 0.214              &  0.214             \\ \hline
Bonn ($\Lambda_{\mu}=1.7$GeV)   & 0.019    & 0.5     & 0.002
& 0.217              &  0.219             \\ \hline\hline
\end{tabular}
\caption{Small-$x$ extrapolation parameters for
         $F_{2p}-F_{2n} (= \alpha x^{\beta})$ and the
         contributions to the Gottfried sum from different
         $x$-regions.}
\end{table}

As a fraction of the total $F_{2D}(x)$ \cite{nmc}, the shadowing correction
amounts to (0.5-1.0\%, 0.4-0.8\%, 0.0-0.3\%) at $x = (0.004, 0.01, 0.1)$,
while the antishadowing is less than 0.2\% of $F_{2D}$ at $x \approx 0.2$.

In fig.19 we show the ratio of neutron structure functions with and without
shadowing corrections,
\begin{eqnarray}
\frac{ F_{2n} }{ (F_{2n})_{NMC} }
& = & 1 - \frac{ \delta F_{2D} }{ F_{2D} }
          \left( \frac{ 1 + (F_{2n}/F_{2p})_{NMC} }{ (F_{2n}/F_{2p})_{NMC} }
          \right)
\end{eqnarray}
where the NMC neutron/proton ratio was defined as
$(F_{2n}/F_{2p})_{NMC} \equiv F_{2D}/F_{2p} - 1$.
There is an overall $1 - 2 \%$ increase in the neutron structure function
due to shadowing for $x \stackrel{<}{\sim} 0.01$.

Finally, we illustrate in fig.20 the dependence upon $Q^{2}$ of the
total shadowing correction, $\delta F_{2D}(x,Q^{2})$.
As expected, the VMD term vanishes rapidly with increasing $Q^{2}$,
leaving the two scaling contributions from ${\cal P}$ and meson exchange
to largely cancel each other for $Q^{2} \simeq 25$ GeV$^{2}$.
However, we should add a note of caution about comparing
shadowing corrections at very large values of $Q^{2}$.
In the parton recombination model \cite{nz75,muq,prec} the fusion of
quarks and gluons from different nucleons introduces additional terms
\cite{muq} in the Altarelli-Parisi equations governing the QCD evolution
of the parton distributions.
At very small $x$ and large $Q^{2}$, such as those attainable at HERA energies,
this can lead to significant corrections \cite{bdkw} to
the $\delta F_{2D}(x,Q^{2})$ evolved without these terms, although the exact
magnitude of these is sensitive to the small-$x$ behaviour of the input
nucleon gluon distribution.
For the moderate range of $Q^{2}$ and not too low $x$ values in
fig.20, however, we expect the indicated $Q^{2}$ behaviour to be
reliable.

\section{Conclusion}

In summary, we have estimated the nuclear shadowing in
lepton-deuteron DIS from the double scattering mechanism in fig.3.
Our approach is similar to that of refs.\cite{nizo} and \cite{bdkw},
in describing the interaction in terms of the VMD model, together with
Pomeron (${\cal P}$) exchange at larger $M_{X}$. However we have also included
contributions from the exchange of mesons which effectively cancel as
much as half of the shadowing from the VMD/${\cal P}$-exchange mechanisms
alone.
Numerically, there is some dependence on the model deuteron wavefunction,
and also on the meson--nucleon form factor for the meson-exchange process.
The net effect is a $\stackrel{<}{\sim} 1\%$ reduction of $F_{2D}$ for
$x \sim 0.004$,
or equivalently a $\stackrel{<}{\sim} 2\%$ increase in the neutron structure
function
over the uncorrected $F_{2n}$.
Consequently, the shadowing correction to the Gottfried sum
$S_{G}(0,1)$ is between --0.010 and --0.026 (or about 4 and 10\% of the
NMC value), which is about 5 times smaller than in previous estimates.

To accurately test the descriptions of shadowing in the deuteron it is
necessary to obtain model-independent information on the neutron
structure function at low $x$.
Even at HERA energies this is not possible with electron scattering alone.
However, when combined with high-precision data from neutrino-proton
experiments the individual flavour distributions can be determined,
and the neutron structure function inferred from charge symmetry.
For this to happen, however, the statistics on the neutrino data need to
be improved, and the range extended to smaller $x$.

\vspace*{1cm}

\hspace*{-0.5cm}
{\large {\bf Acknowledgements}}

\hspace*{-0.5cm}
W.M. would like to thank M.Gari, J.Haidenbauer, B.Loiseau and N.Nikolaev
for helpful discussions.
This work was supported by the Australian Research Council.

\newpage


\newpage

{\large {\bf Figure captions}}

\vspace*{0.5cm}

\normalsize

{\bf 1.} Lepton-deuteron deep inelastic scattering.

{\bf 2.} Virtual photon-deuteron scattering in the impulse approximation.

{\bf 3.} Double scattering of virtual photons from the deuteron,
         in which both nucleons take part in the interaction.

{\bf 4.} Double scattering mechanism in the vector meson dominance model.
         The virtual photon dissociates into a vector meson which then
         scatters from the nucleon.

{\bf 5.} VMD contribution to the total deuteron structure function at
         $Q^{2} = 4$ GeV$^{2}$, for different model deuteron form factors.

{\bf 6.} Deuteron form factor, as defined by eqn.(\ref{SD}).

{\bf 7.} Pomeron exchange contribution to the double scattering mechanism
         (the Pomeron is denoted by the zig-zag).

{\bf 8.} Exchange-Pomeron distribution function, $|f_{{\cal P}}(y)|$.

{\bf 9(a).} Quark-antiquark box contribution to the Pomeron structure function,
            where the Pomeron couples to the virtual photon via a
            quark-antiquark pair.

{\bf\ (b).} Triple-Pomeron contribution to the Pomeron structure function.

{\bf 10.} $x_{{\cal P}}$ dependence of the `box' and $3{\cal P}$ contributions
          to $F_{2{\cal P}}(x_{{\cal P}})$ for the quark distribution function
          parameterisations of Owens \cite{ow} and Morfin and Tung \cite{mt}
          at $Q^{2} = 4$ GeV$^{2}$.

{\bf 11(a).} Quark-antiquark box and $3{\cal P}$ contributions to the total
             deuteron structure function. The deuteron form factor is given
             by the Bochum model wavefunction.

{\bf\ \ (b).} Deuteron form factor dependence of the Pomeron exchange
              contribution to the deuteron structure function.

{\bf 12.} Double scattering mechanism with meson exchange.
          The dotted line represents mesons $\pi, \rho, \omega, \sigma$.

{\bf 13.} Individual meson exchange contributions to the deuteron structure
          function, for the wavefunction of the Bonn (OBEPQ) model with
          a universal form factor cut-off $\Lambda_{\mu} = 1.7$ GeV.
          Note the mass of the effective $\sigma$ meson is $\approx$ 800 MeV.

{\bf 14.} Deuteron wavefunction and $\mu NN$ form factor dependence
          of the total meson exchange correction.
          The Bochum (solid) and Paris (dashed) curves are evaluated
          with $\Lambda_{\mu} = 1.3$ GeV, while the Bonn (dotted)
          curves have $\Lambda_{\mu} = 1.0, 1.3$ and 1.7 GeV,
          with the larger cut-off giving more overall antishadowing.

{\bf 15.} Comparison between the VMD, Pomeron and meson exchange corrections
          to the deuteron structure function at $Q^{2} = 4$ GeV$^{2}$
          (for the Bochum wavefunction, and a form factor cut-off
          $\Lambda_{\mu} = 1.3$ GeV for the meson exchange process).

{\bf 16.} Deuteron wavefunction and $\mu NN$ form factor dependence of the
          total shadowing correction. For the Bochum and Paris curves
          $\Lambda_{\mu} = 1.3$ GeV, while the Bonn curves are calculated
          with $\Lambda_{\mu} = 1.0, 1.3$ and 1.7 GeV,
          with the larger cut-off giving less overall shadowing.

{\bf 17.} Difference between the proton and neutron structure functions,
          with shadowing corrections to the NMC data at $Q^{2} = 4$ GeV$^{2}$.

{\bf 18.} Gottfried sum, with shadowing corrections to the NMC data.

{\bf 19.} Neutron structure function ratio, with and without shadowing
          corrections (at $Q^{2} = 4$ GeV$^{2}$).

{\bf 20.} $Q^{2}$ dependence of the total shadowing correction to $F_{2D}$.
          Curves represent the shadowing corrections at 4, 10 and 25
          GeV$^{2}$, with the Bochum model wavefunction and with
          $\Lambda = 1.3$ GeV. Also shown is the correction at
          $Q^{2} = 25$ GeV$^{2}$ without the VMD contribution.
\end{document}